\documentclass{ws-procs9x6}

\begin{document}

\title{Neutrino Processes in Supernovae and Neutrons Stars in Their 
Infancy and OLd Age}

\author{M. Prakash, S. Ratkovi\'c \& S. I. Dutta}
\address{Department of Physics \& Astronomy \\ 
State University of New York at Stony Brook \\
Stony Brook, NY 11794-3800, USA}  
\author{PALS\footnote{\uppercase{F}riends who have helped
and contributed significantly; see the
\uppercase{A}cknowledgements.}}


\maketitle

\abstracts{Neutrino processes in semi-transparent supernova matter,
opaque to semi-transparent protoneutron star matter, and catalyzed
neutron stars  are discussed.}

\section{Introduction}

In recent years, the study of neutrino emission, scattering, and
absorption in matter at high density and/or temperature has gained
prominence largely due to its importance in a wide range of
astrophysical phenomena.  Energy loss in degenerate helium cores of
red giant stars,\cite{Gross,Raffelt(2000)} cooling in pre-white dwarf
interiors,\cite{O'BrienKawaler(2000)} the short- and long-term cooling
of neutron stars,\cite{Prak02,Yak02} the deflagration stages of white
dwarfs which may lead to type Ia supernovae,\cite{Wolfgang,Iwamoto}
explosive stages of type II (core-collapse) supernovae,\cite{Bur00}
and thermal emission in accretion disks of gamma-ray
bursters,\cite{Matteo,Kohri} are examples in which neutral and charged
current weak interaction processes that involve neutrinos play a
significant role.  Since this is a vast subject, we will highlight
some recent developments in the context of core-collapse supernovae
and neutron stars from their birth to old age.

\section{Supernovae}

In unravelling the mechanism by which a type-II supernova explodes,
the implementation of accurate neutrino transport has been realized to
be critical.\cite{Trans}  The basic microphysical inputs of accurate
neutrino transport coupled in hydrodynamical situations are the
differential neutrino production and absorption rates and their
associated emissivities.  The processes and precise forms in which
such inputs are required for multienergy treatment of neutrinos for
both sub-nuclear and super-nuclear densities (nuclear density $\rho_0
\simeq 2.65 \times 10^{14}~{\rm g~cm^{-3}})$ are detailed in
Refs. [\refcite{BRUENN1,BT02}]. 

\begin{figure}[ht]
\centerline{\epsfxsize=3.7in\epsfbox{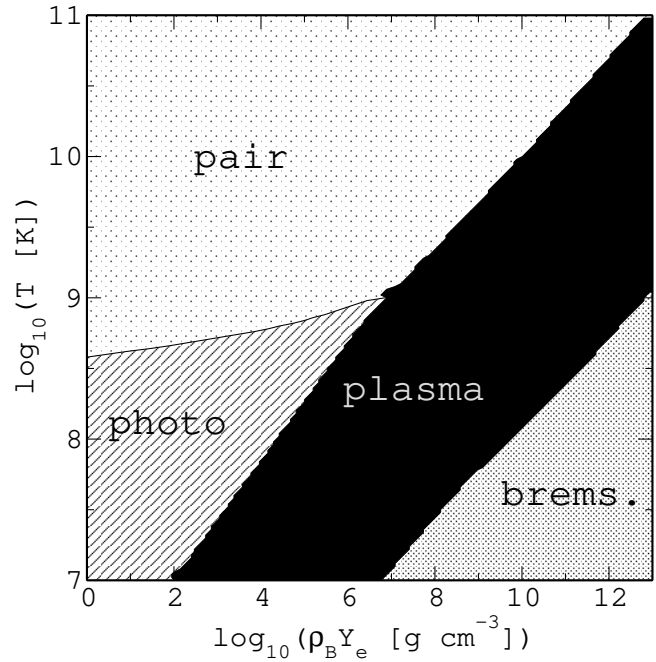}}   
\caption{Regions of temperature and density in which the various
neutrino emitting processes are operative.  }
\label{fig:comparenu}
\end{figure}

At sub-nuclear densities, the relevant processes are:
\begin{eqnarray}
\label{pair1}
{\rm pair~production}~&:&~
e^+ + e^- \rightarrow \nu + \bar\nu\,, \\
\label{photo}
{\rm the~photo-neutrino~process}~&:&~
e^\pm + \gamma^* \rightarrow e^\pm + \nu + \bar\nu\,,  \\
\label{plasma}
{\rm the~plasma~process}~&:& 
\gamma^* \rightarrow \nu +\bar\nu\,,  \\
\label{brems}
{\rm nucleon-nucleon~bremsstrahlung}~&:&~
(n,p) \rightarrow (n,p)+ \nu + \bar\nu\,,~{\rm and}  \\
\label{flavor}
\nu-{\rm flavor~production}~&:& 
\nu_i + \bar{\nu}_i \rightarrow \nu_j + \bar{\nu}_j\,. 
\end{eqnarray}
The relative importance of these processes depends on the temperature
and density of ambient matter and is sketched in
Figure~\ref{fig:comparenu}.

Additional neutrino processes in the supernova environment include
\begin{equation}
p + e^- \rightarrow n + \nu_e,\quad (A,Z) + e^- \rightarrow (A,Z-1) + 
\nu_e\,,  
\label{neu}
\end{equation}
\begin{equation}
\nu + (A,Z) \rightarrow \nu + (A,Z) \,,  
\label{coh}
\end{equation}
\begin{equation}
\nu + e^- \rightarrow \nu + e^-, \qquad
\nu + (A,Z) \rightarrow \nu + (A,Z)^*\,, 
\label{scat}
\end{equation}
\begin{eqnarray}
(A,Z)^* \rightarrow  (A,Z) + \nu + \bar{\nu}\,. 
\label{pair2}
\end{eqnarray}
Reactions (\ref{neu}) begin the process of neutronization and decrease
of lepton number per baryon $Y_L$, whose value after $\nu$-trapping
determines the masses of the homologous core and initial PNS, and thus
the available energy for the shock and subsequent neutrino emissions.
The equation of state also influences these quantities, most
importantly through the nuclear symmetry energy.  In the subnuclear
density regime, the coherent scattering reaction (\ref{coh}) from
nuclei in a lattice is the most important opacity source.  The
reactions (\ref{scat}) are important in changing the neutrino energy
and in achieving thermodynamic equilibrium.  The reactions
(\ref{pair1}) and (\ref{pair2}) are also important in achieving
thermodynamic equilibrium.  The bremsstrahlung and modified Urca
($n+p\rightarrow n+n+e^++\nu+\bar\nu$) processes dominate in many
circumstances.  For example, the production and thermalization of
$\mu$ and $\tau$ neutrinos, which receives contributions from
reactions (\ref{pair1}) through (\ref{flavor}), and (\ref{pair2}), is
dominated by nucleon bremsstrahlung (\ref{brems}) for number density
$n>0.005~{\rm fm}^{-3}$ and temperature $T<15$ MeV.  The modified Urca
process dominates the cooling of protoneutron stars if direct Urca
processes involving nucleons, hyperons or other strange particles do
not occur.

\section{Kernels for Neutrino Transport Calculations}
\label{sec:PKER}

The evolution of the neutrino distribution function $f$, generally
described by the Boltzmann transport equation in conjunction with
hydrodynamical equations of motion together with baryon and lepton
number conservation equations, is 
\begin{eqnarray}
\label{BOLTZMANN}
\frac{\partial f}{\partial t}+v^i\frac{\partial f}{\partial x^i}+
\frac{\partial (f F^i)}{\partial p^i}
=B_{EA}(f)+B_{NES}(f)+B_{\nu{ N}}(f)+B_{TP}(f) \,.
\end{eqnarray}
Here, $F^i$ is the force acting on the particle and we have ignored
general relativistic effects for simplicity (see, for example,
Ref. [\refcite{BRUENN1}] for full details).  The right hand side of the
above equation is the neutrino source term in which, $B_{EA}(f)$
incorporates neutrino emission and absorption processes, $B_{NES}(f)$
accounts for the neutrino-electron scattering process, $B_{\nu{
N}}(f)$ includes scattering of neutrinos off nucleons and nuclei, and
$B_{TP}(f)$ considers the thermal production and absorption of
neutrino-antineutrino pairs.

Till recently, detailed differential information was not available for
the plasma and photoneutrino processes. In prior works in which the
total rates and emissivities for these processes were computed, the
energy and angular dependences of the emitted neutrinos with 4-momenta
$q$ and $q^{\prime}$ were eliminated by using Lenard's identity:
\begin{equation}
\int \frac {d^3q}{2E_q}   \frac {d^3q^{\prime}}{2E_{q^{\prime}}} 
\delta^4(q_t-q-q^{\prime} ) q^\mu q{^{\prime\nu}} = \frac {\pi}{24} 
\Theta(q_t^2) (2 q_t^\mu q_t^\nu + q_t^2 g^{\mu\nu})  \,.  
\label{Lenard}
\end{equation}  
Although the use
of this identity simplifies considerably the calculation of the total
emissivity, differential information about the neutrinos is entirely
lost.  On the other hand, calculations of differential rates and
emissivities, such as
\begin{equation}
\frac {d^3\Gamma}{dE_q\, dE_{q^\prime}\, d(\cos\theta_{qq^{\prime}})} \, 
\qquad {\rm and} \qquad 
\frac {d^3Q}{dE_q\, dE_{q^\prime}\, d(\cos\theta_{qq^{\prime}})} \, , 
\label{diffs}
\end{equation} 
where $\theta_{qq^{\prime}}$ is the angle between the neutrino pairs,
entail the calculation of the relevant squared matrix elements hitherto
bypassed in obtaining the total rates and emissivities.  Realizing
this, the squared matrix elements for the plasma and photoneutrino
processes were computed in Refs.~[\refcite{rdp03,drp04}]. Some results
from these recent works are highlighted below.

\subsection{The Plasma Process}

\begin{figure}[ht]
\centerline{\epsfxsize=3.7in\epsfbox{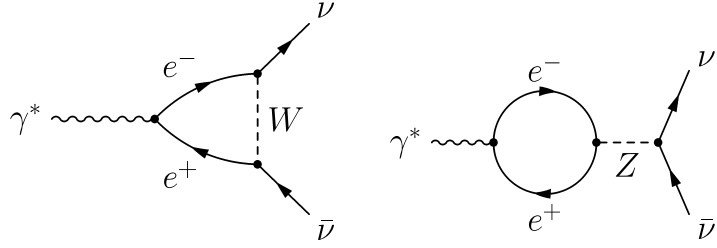}}   
\caption{Leading order Feynman diagrams describing the emission of a
neutrino pair from the plasma process. The charged current process
in which the $W$-boson is exchanged produces only $\nu_e \bar{\nu_e}$,
while that in which the neutral $Z$-boson is exchanged results in pairs of all
three neutrino ($e,~\mu,~{\rm and}~\tau$) flavors.}
\label{fig:fplasma}
\end{figure}

As is well known, $e^+e^-$ pairs in a plasma cause the photon to
acquire an effective mass, which arises from electromagnetic
interactions (cf. Refs. [\refcite{rdp03,BRAATEN1}] and references therein).
Therefore, we can consider the photon to be a massive spin--1 particle
that couples to the $\nu\bar{\nu}$ pair through the two one-loop
diagrams shown in Figure  \ref{fig:fplasma}.

Suppressing the dependencies on $(r,t)$ for notational simplicity, 
the source term for the plasma process can be written as
\begin{eqnarray}
\label{BTP}\nonumber
B\big(f\big) &=& 
[1-f]\frac{1}{(2\pi)^3}
\int_0^\infty E^2_2 \,\,dE_2 \int_{-1}^1 
d\mu_2 \int_0^{2\pi}d\phi_2 \,R^p
(E_1,E_2,cos\,\theta )
\,[1-\bar{f}]\\
&-&  f\frac{1}{(2\pi)^3}
\int_0^\infty E^2_2 \,\,dE_2 \int_{-1}^1 d\mu_2 
\int_0^{2\pi}d\phi_2
\,R^a(E_1,E_2,cos\,\theta )
\,\bar{f} \,,
\end{eqnarray}
where the first and the second terms correspond to the source
(neutrino gain) and sink (neutrino loss) terms, respectively.  $E_1$
and $E_2$ are the energies of the neutrino and antineutrino,
respectively.  Angular variables $\mu_i\equiv \cos\theta_i$ and
$\phi_i$ ($i=1,2$) are defined with respect to the $z$-axis that is
locally set parallel to the outgoing radial vector ${\bf r}$. The
angle $\theta$ between the neutrino and antineutrino pair is related
to $\theta_1$ and $\theta_2$ through
\begin{eqnarray}
\cos\theta &=&
\mu_1\mu_2+\sqrt{(1-\mu_1^2)(1-\mu_2^2)}~\cos(\phi_1-\phi_2) \,.
\end{eqnarray}
Notice that $f \equiv f(E_1,\mu_1)$ and $\bar{f} \equiv \bar{f}(E_2,\mu_2)$. 
The production and absorption kernels are given by
\begin{eqnarray}\label{kernel}
R^{\genfrac{}{}{0pt}{}{p}{a}}(E_1, E_2, \cos\theta)&=& 
\int \frac{d^3k}{(2\pi)^3}\, Z_Y(k)\, 
\genfrac{(}{)}{0pt}{}{\xi n_B(\omega, T)}{1+n_B(\omega, T)} 
\nonumber \\
&& \times \frac{1}{8\omega E_1 E_2}  \,
\delta^4(K-Q_1-Q_2)(2\pi)^4\langle {|\mathcal{M}|}^2\rangle\,, 
\end{eqnarray}
where the subscript $Y$ stands for $T$--``transverse'' or
$L$--``longitudinal''. The factor $\xi$ accounts for the spin
avaraging; $\xi=2$ for the transverse, axial and mixed cases, while
for the longidudinal case $\xi=1$.

The angular dependences in the kernels $R^{\genfrac{}{}{0pt}{}{p}{a}}(E_1, E_2,
\cos\theta)$ are often expressed in terms of Legendre polynomials as
\begin{eqnarray}\label{RLegendre}
R^{\genfrac{}{}{0pt}{}{p}{a}}(E_1, E_2, \cos\theta)&=&
\sum_{l=0}^{\infty} \frac{2l+1}{2} 
\Phi^{\genfrac{}{}{0pt}{}{p}{a}}_l(E_1, E_2) 
P_l(\cos\theta) \,, 
\end{eqnarray}
where the Legendre coefficients
$\Phi^{\genfrac{}{}{0pt}{}{p}{a}}_l(E_1, E_2)$ depend exclusively on
energies.
 
From Eq.~(\ref{kernel}), it is evident that the kernels are related
to the neutrino rates and emissivities.  We first consider the
production kernel $R^{p}(E_1, E_2, \cos\theta)$.  The corresponding
analysis for the absorption kernel $R^{a}(E_1, E_2, \cos\theta)$ can
be made along the same lines, but with the difference that $n_B$ is
replaced by $1+n_B$.  The neutrino production rate is given by
\begin{eqnarray}\label{rate}
\Gamma&=&\xi\,\int \frac{d^3k}{(2\pi)^3 2\omega}\, Z_Y(k)\, 
\frac{d^3q_1}{(2\pi)^3 2E_1}
\frac{d^3q_2}{(2\pi)^3 2E_2} \nonumber \\
&& \times \, n_B(\omega,T) \, \delta^4(K-Q_1-Q_2)(2\pi)^4
\langle{|\mathcal{M}|}^2\rangle \nonumber \\
&=& \int \frac{d^3q_1}{(2\pi)^3} \frac{d^3q_2}{(2\pi)^3} 
R^{p}(E_1, E_2,\cos\theta)\,, 
\end{eqnarray}
which defines the kernel $R^p(E_1, E_2,\cos\theta)$ and is to be
identified with that in Eq.~(\ref{kernel}).
The emissivity $Q$ can also be cast in terms of $R^p$ using 
\begin{equation}
Q=\int \frac{d^3q_1}{(2\pi)^3} \frac{d^3q_2}{(2\pi)^3} (E_1+E_2) R^{p}(E_1,
E_2, \cos\theta)\,. \label{Remiss}
\end{equation}

\begin{figure}[h!]
\centerline{\epsfxsize=4.7in\epsfbox{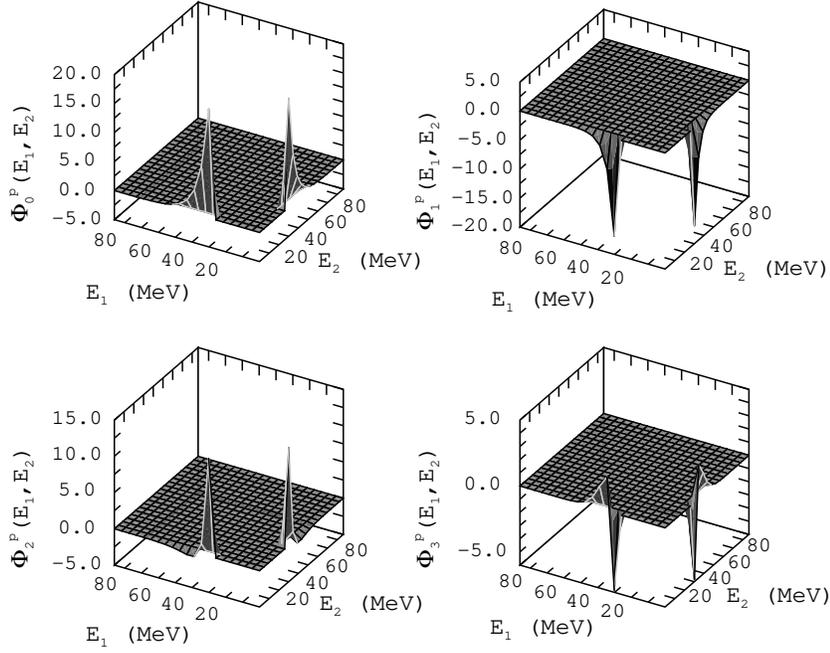}}   
\caption{The transverse part of the production kernels for $T=8.62$
MeV and $\rho_B Y_e<10^{10}$ g cm$^{-3}$. The Legendre coefficients
$\Phi_l^{p}$ are shown for $l=0$ through $3$. The
neutrino energies $E_1$ and $E_2$ are in MeV and the Legendre
coefficients are in units of $10^{129}$ $\hbar^6$ erg$^{-6}$ cm$^3$
s$^{-7}$. }
\label{fig:PHIS_plasma}
\end{figure}

\begin{figure}[ht]
\centerline{\epsfxsize=3.7in\epsfbox{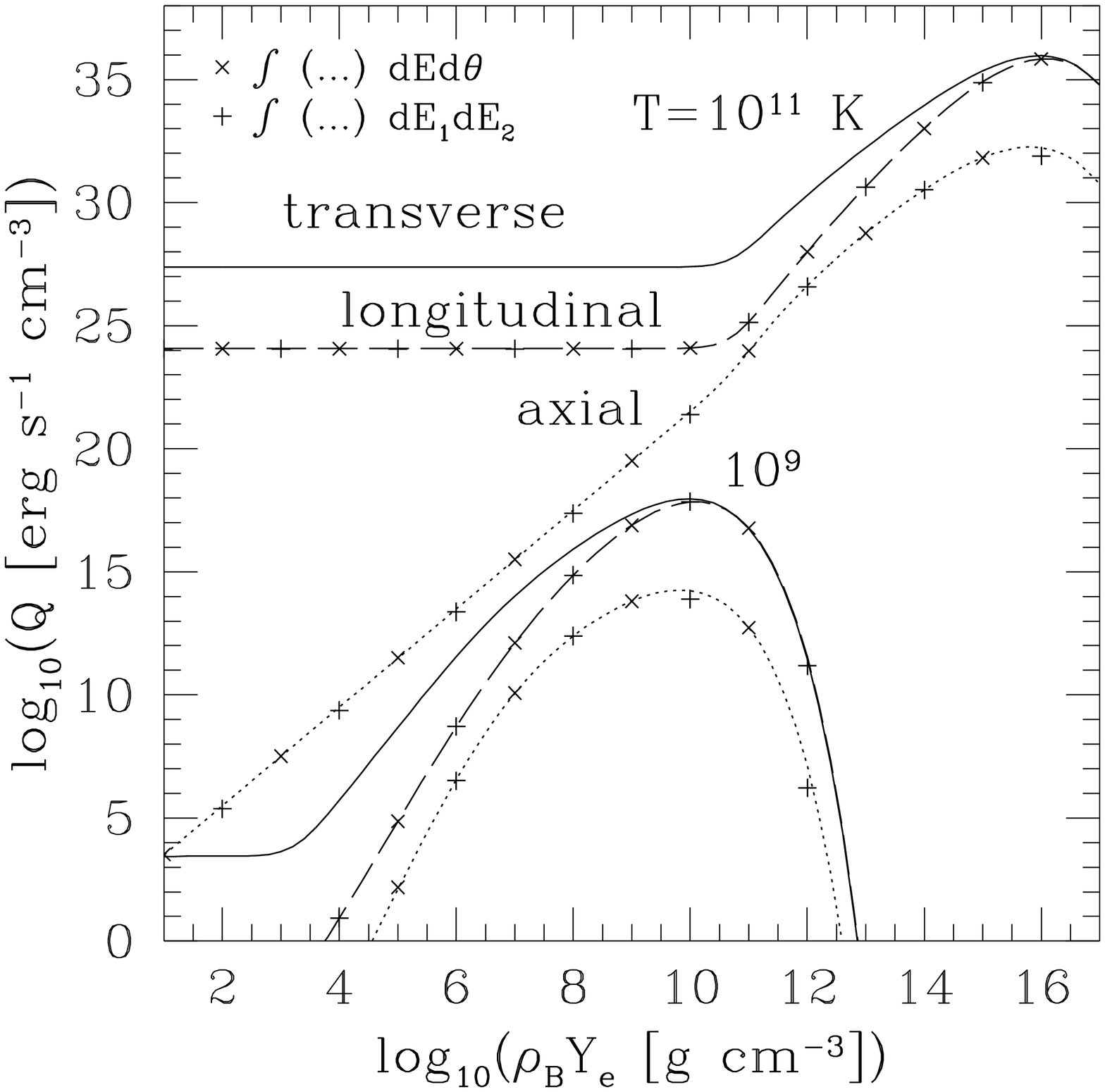}}   
\caption{Individual contributions from the transverse, longitudinal,
and axial channels to the neutrino emissivity.  The mass density of
protons in the plasma, $\rho_BY_e = m_pn_e$, where $m_p$ is the proton
mass, $Y_e = n_e/n_B$ is the net electron fraction ($n_B$ is the
baryon number density) , and $n_e$ is net electron number density.}
\label{fig:QTLA}
\end{figure}

The Legendre coefficients $\Phi_l^p$ for the transverse part of the
production kernels are shown in Figure \ref{fig:PHIS_plasma} for
$l=0$ through 3. The longitudinal component becomes comparable to
the transverse component only in the strongly degenerate regime (see
Figure \ref{fig:QTLA}), while the axial and mixed components (the
latter contributes only to the differential rates and emissivities but
not to the total, since it is anti-symmteric in $E_1$ and $E_2$) are
negligibly small. Notice that the first few Legendre coefficients are
all comparable in magnitude. Moreover, the emission process is
anisotropic (see also Figure 7 of Ref. \refcite{rdp03}).  ``Light'' photon
($\omega_p \ll T$) decays result in neutrino pairs with small outgoing
angles between them, whereas ``massive'' photons ($\omega_p \gg T$)
yield back-to-back neutrino emission.

Figure \ref{fig:QTLA} shows the individual contributions to the total
emissivity from the transverse, longitudinal, and axial channels at
$T=10^{11}$K and $T=10^9$K, respectively.  The curves show results
from expressions derived by exploiting the Lenard
identity.\cite{rdp03,BRAATEN1} The symbols ``$\times$'' and ``$+$''
show results obtained by integrations of the differential
emissivities.  At all densities and temperatures, the contribution of
the axial channel to the total emissivity is negligible.  For each
temperature, the emissivity in the transverse channel dominates over
that of the longitudinal channel at low densities. However, the peak
values in these two channels are attained at nearly the same density;
thereafter their individual contributions coincide.

In terms of the density and temperature dependencies of the plasma
frequency $\omega_p$ and the electron chemical potential $\mu_e$ (see
Figure~4 of Ref.~[\refcite{rdp03}]), a detailed qualitative and quantitative
analyses of the basic features of Figure~\ref{fig:QTLA} are provided in
Ref.~[\refcite{rdp03}].  Consider the case of $Q_T$ first.  At a fixed
temperature, the basic features to note are: \\

\noindent (1) $Q_T$ is independent of the density
$\rho_BY_e$ until a turn-on density $(\rho_BY_e)_{\rm to}$ is reached,
\\
\noindent (2) For densities larger than this turn-on density, $Q_T$ exhibits a
power-law rise until a maximum is reached at $(\rho_BY_e)_{\rm peak}$, and \\
\noindent (3) For $\rho_BY_e \gg (\rho_BY_e)_{\rm peak}$, the fall-off
with density is exponential.

In the case of $T \gg \omega_p$ (non-degenerate plasma), the main
contribution to $Q_T$ comes from high photon momenta.  To a very good
approximation,\cite{BRAATEN1}
\begin{equation}\label{QTAT}
Q_T \simeq \frac{2\sum_f {(C_V^f)}^2 
G_F^2}{48 \pi^4 \alpha} \zeta(3) m_T^6 T^3
\,,
\end{equation}
where $\zeta(3) \simeq 1.202$ is Riemann's Zeta function and $m_T$ is
the transverse photon mass.

For $T \ll \omega_p$ (degenerate plasma), $Q_T$ takes the form\cite{BRAATEN1} 
\begin{equation}\label{QTAW}
        Q_T \simeq \frac{\sum_f {(C_V^f)}^2 G_F^2}{48 \pi^4 \alpha} 
\sqrt{\frac{\pi}{2}}  \omega_p^{15/2} T^{3/2} e^{-\omega_p/T} \,.
\end{equation}
The analyses of the longitudinal and axial 
emissivities can be carried out along the same lines as that for
the transverse emissivity.\cite{rdp03}

\subsection{The Photoneutrino Process} 

\begin{figure}[ht]
\centerline{\epsfxsize=3.2in\epsfbox{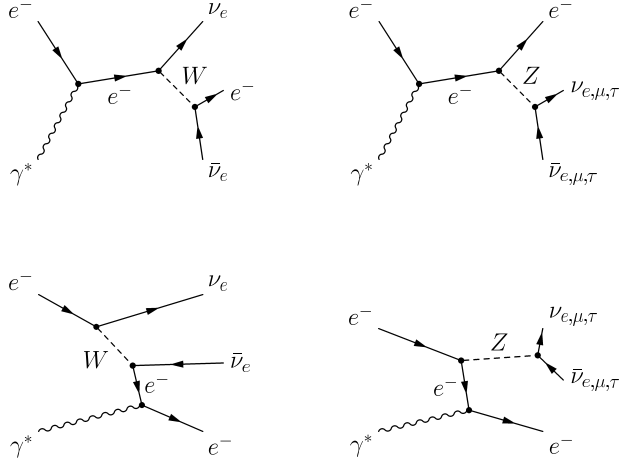}}   
\caption{Leading order Feynman diagrams describing the emission of a
neutrino pair from the photoneutrino process.  The charged current $W$ -
exchange channel produces only $\nu_e \bar{\nu}_e$, whereas the
neutral $Z$ - exchange results in pairs of all three neutrino
($e,~\mu,~{\rm and}~\tau$) flavors. Contributions from positrons are
obtained by the replacement $e^- \rightarrow e^+$.}
\label{fig:fphoto}
\end{figure}

The leading order diagrams for the photoproduction of neutrino pairs,
$e^{\pm} + \gamma^* \rightarrow e^{\pm} + \nu_{e,\mu,\tau} +
\bar\nu_{e,\mu,\tau}$, are shown in Figure~\ref{fig:fphoto}.  The kernels of
the collision integral for the photoneutrino process can be
calculated along the same lines as for the plasma process and are
detailed in Ref. \refcite{drp04}.

\begin{figure}[ht]
\centerline{\epsfxsize=5.2in\epsfbox{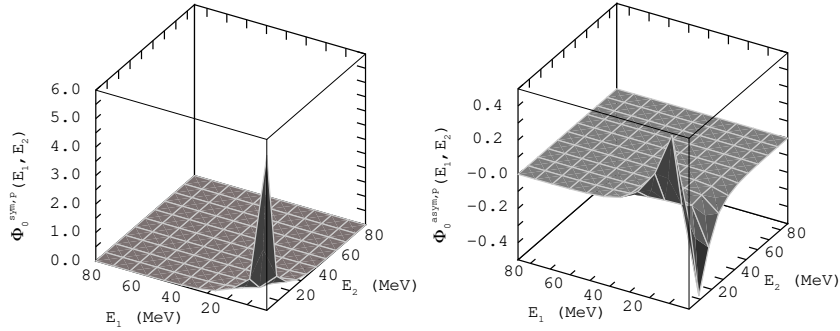}}   
\caption{Symmetric and anti-symmetric parts of the Legendre
coefficients $\Phi_0^{p\,({\rm sym})}$ and $\Phi_0^{p\,({\rm asym})}$
in the production kernels for the
transverse case.  The $\nu$-energies $E_q$ and
$E_{q^\prime}$ are in MeV and the Legendre coefficients 
are in units of $10^{129}$ $\hbar^6$ erg$^{-6}$ cm$^3$ s$^{-7}$.}
\label{fig:PHIS_photo}
\end{figure}

Numerical results for the symmetric and anti-symmetric components of
$\Phi_0^p(E_q,E_{q^\prime})$ and $\Phi_1^p(E_q,E_{q^\prime})$ are
shown in Figure \ref{fig:PHIS_photo} for the case $T=10^{11}$ K$=8.62$
MeV and $\rho_BY_e=1$ g cm$^{-3}$.  The results explicitly show the
expected symmetry properties in neutrino energies $E_q$ and $E_{q^\prime}$.  A
comparison of the relative magnitudes $\Phi_1^p(E_q,E_{q^\prime})$
(not shown here, but see Figure~6 of Ref.~[\refcite{drp04}]) and $\Phi_0^{p\,
({\rm sym})}(E_q,E_{q^\prime})$ shows that the $l=0$ term is the
dominant term.  The magnitude of $\Phi_0^{p\, ({\rm asym})}$ amounts
to only 10\% of the leading $\Phi_0^{p\, ({\rm sym})}$
contribution. The contributions of $\Phi_1^{p\, ({\rm sym})}$ and
$\Phi_1^{p\, ({\rm asym})}$ are 6\% and 3\%, respectively.  In
physical terms, this means that neutrino-pair emission from the
photo-neutrino process is dominantly isotropic.  Therefore, depending
on the required accuracy, $\Phi_0^{p\, ({\rm sym})}$ might be adequate
in practical applications.  Note also that that the production kernels
are negligible for energies $E_q$ and $E_{q^\prime}$ $\gtrsim 10T$.

\begin{figure}[ht]
{\epsfxsize=4in\epsfbox{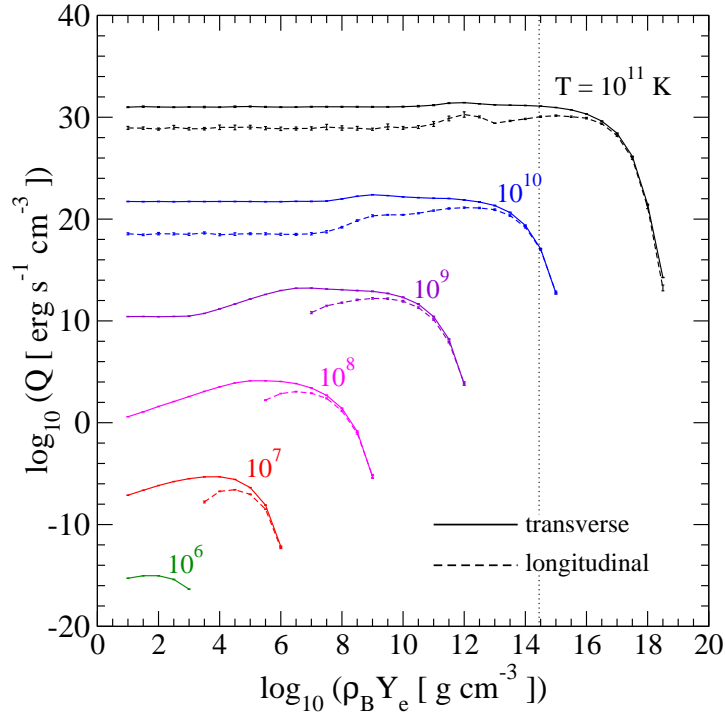}}   
\caption{Individual contributions from the transverse and longitudinal
channels to the neutrino emissivity as a function of baryon density at
the indicated temperatures.  The error bars show the variance of the
Monte Carlo integration.  For densities in excess of nuclear density
shown by the dotted vertical line, neutrino production from strongly
interacting particles dominate over QED-plasma processes. }
\label{fig:qphoto}
\end{figure}

Contributions of the transverse and longitudinal components to the
total emissivity are shown in Figure \ref{fig:qphoto}. For all
temperatures at sufficiently high net electron densities $n_e$ (i.e.,
a degenerate plasma), the inequalities
\begin{eqnarray}
\mu_e \gg T,&\;& \mu_e \gg m_e,\; \nonumber \\ 
\omega_P \gg T,&\;& \omega_P \gg m_e
\end{eqnarray}
are satisfied. In this  case, 
\begin{eqnarray} \label{eq:qtdeg}
	Q_T&\simeq&\frac{4}{3} \, \frac{\alpha G_F^2
	(C_V^2+C_A^2)}{(2\pi)^{6}}\,
	\, {\omega_p}^6 \, T^3 \, e^{-\omega_p/T}\, .
\end{eqnarray}

The nondegenerate situation occurs at sufficiently low densities for
which $\mu_e-m_e \ll T$.  In this case, both $Q_T$ and $Q_L$ exhibit a
plateau for temperatures $T \ge 10^9$ K. 
The emissivity and rate can be expressed in terms of the simple expressions
\begin{eqnarray}
  \genfrac{(}{)}{0pt}{}{Q_T}{\Gamma_T}&=&
  \frac{20 \alpha G_F^2 (C_V^2+C_A^2)}{3(2\pi)^{6}}
  \, \genfrac{(}{)}{0pt}{}{T^9\times 775.54 }{T^8\times 136.50}\,.
\end{eqnarray}

The significance of the detailed differential rates and emissivities
of neutrino emission and absorption processes in calculations of
core-collapse supernovae in which neutrino transport is strongly
coupled with hydrodynamics remains to be explored.

\section{Protoneutron Stars}

A protoneutron star (PNS) is born in the aftermath of the
gravitational collapse of the core of a massive star accompanying a
successful supernova explosion.  During the first tens of seconds of
evolution, nearly all ($\sim$ 99\%) of the remnant's binding energy is
radiated away in neutrinos of all
flavors.\cite{burr86,KJ95,burr99a,pons99,Pon00a,Pon01b} The neutrino
luminosities and the emission timescale are controlled by several
factors, such as the total mass of the PNS and the opacity at
supranuclear density, which depends on the composition and dense
matter equation of state (EOS).  One of the chief objectives in
modeling PNS's is to infer their internal compositions from neutrino
signals detected from future supernovae by SuperK, SNO and others
under consideration, including UNO.\cite{AIP}

\subsection{The Evolution of a Protoneutron Star}

\begin{figure}[ht]
\centerline{\epsfxsize=4.7in\epsfbox{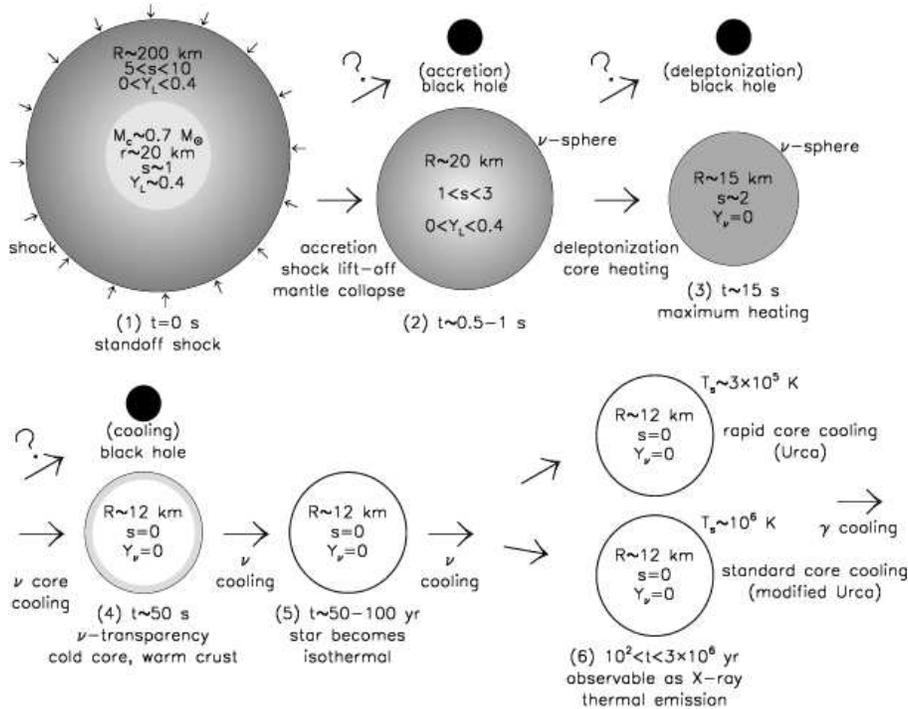}}   
\caption{The main stages of evolution of a protoneutron star. Shading
approximately indicates relative temperatures. }
\label{fig:pns}
\end{figure}

The evolution of a PNS proceeds through several distinct
stages\cite{burr86,supernova} and with various outcomes,\cite{prak97a}
as shown schematically in Figure~\ref{fig:pns}. Immediately following
core bounce and the passage of a shock through the outer PNS's mantle,
the star contains an unshocked, low entropy core of mass $\simeq0.7$
M$_\odot$ in which neutrinos are trapped (stage 1 in the figure). The
core is surrounded by a low density, high entropy ($5<s<10$) mantle
that is both accreting matter from the outer iron core falling through
the shock and also rapidly losing energy due to electron captures and
thermal neutrino emission. The mantle extends up to the shock, which
is temporarily stalled about 200 km from the center prior to an
eventual explosion.

After a few seconds (stage 2), accretion becomes less important if the
supernova is successful and the shock has ejected the stellar envelope.
Extensive neutrino losses and deleptonization will have led to a loss
of lepton pressure and the collapse of the mantle.  If enough
accretion has occurred, however, the star's mass could increase beyond the
maximum mass capable of being supported by the hot, lepton-rich
matter.  If this occurs, the remnant collapses to form a black hole
and its neutrino emission is believed to quickly cease.\cite{burr88}

Neutrino diffusion deleptonizes the core on time scales of 10--15 s
(stage 3).  Diffusion time scales are proportional to
$R^2(c\lambda_\nu)^{-1}$, where $R$ is the star's radius and
$\lambda_\nu$ is the effective neutrino mean free path.  This generic
relation illustrates how both the EOS and the composition influence
evolutionary time scales.  The diffusion of high-energy (200--300 MeV)
$\nu$'s from the core to the surface where they escape as low-energy
(10--20 MeV) $\nu$'s generates heat (a process akin to joule
heating). The core's entropy approximately doubles, producing
temperatures in the range of 30--60 MeV during this time, even as
neutrinos continue to be prodiguously emitted from the star's effective
surface, or $\nu-$sphere.

Strange matter, in the form of hyperons, a Bose condensate, or quark
matter, suppressed when neutrinos are trapped, could appear
at the end of the deleptonization.  Its appearance would
lead to a decrease in the maximum mass that matter is capable of
supporting, implying metastability of the neutron star and another
chance for black hole formation.\cite{prak97a}  This would occur
if the PNS's mass, which must be less than the maximum mass of hot,
lepton-rich matter (or else a black hole would already have formed),
is greater than the maximum mass of hot, lepton-poor matter.  However,
if strangeness does not appear, the maximum mass instead increases
during deleptonization and the appearance of a black hole would be
unlikely unless accretion in this stage remains significant.

The PNS is now lepton-poor, but it is still hot.  While the star has
zero net neutrino number, thermally produced neutrino pairs of all
flavors dominate the emission.  The average neutrino energy slowly
decreases, and the neutrino mean free path increases.  After
approximately 50 seconds (stage 4), $\lambda\simeq R$, and the star
finally becomes transparent to neutrinos.  Since the threshold density
for the appearance of strange matter decreases with decreasing
temperature, a delayed collapse to a black hole is still possible
during this epoch.

Following the onset of neutrino transparency, the core continues to
cool by neutrino emission, but the star's crust remains warm and cools
less quickly. The crust is an insulating blanket which prevents the
star from coming to complete thermal equilibrium and keeps the surface
relatively warm ($T\approx3\times10^6$ K) for up to 100 years (stage
5).  The temperature of the surface after the interior of the
star becomes isothermal (stage 6) is determined by the rate of
neutrino emission in the star's core and the composition of the surface.

\subsection{Neutrino Signals in Terrestrial Detectors}

\begin{figure}[ht]
\centerline{\epsfxsize=3.7in\epsfysize=4in\epsfbox{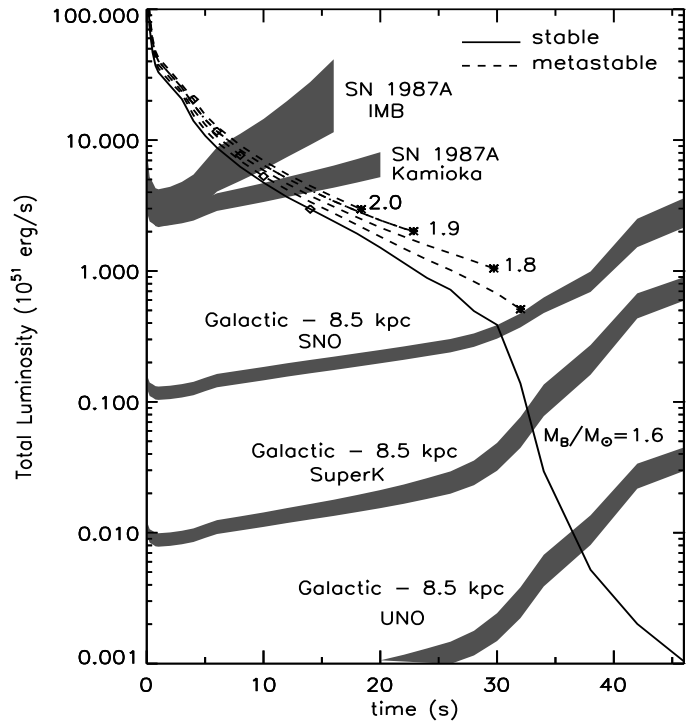}}   
\caption{The evolution of the total neutrino luminosity for $npQ$
PNS's.  Shaded bands illustrate the limiting luminosities corresponding
to a count rate of 0.2 Hz, assuming a supernova distance of 50 kpc for
IMB and Kamioka, and 8.5 kpc for SNO and SuperK. The widths of the
shaded regions represent uncertainties in the average neutrino energy
from the use of a diffusion scheme for neutrino transport.}
\label{fig:lum1}
\end{figure}

A comparison of the signals observable with different detectors is
shown in Figure~\ref{fig:lum1}, which displays $L_\nu$ as a function
of baryon mass $M_B$ for stars containing quarks in their cores.  In
the absence of accretion, $M_B$ remains constant during the evolution,
while the gravitational mass $M_G$ decreases.  The two upper shaded
bands correspond to estimated SN 1987A (50 kpc distance) detection
limits with KII and IMB, and the lower bands correspond to estimated
detection limits in SNO, SuperK, and UNO, for a Galactic supernova
(8.5 kpc distance).  The detection limits have been set to a count
rate $dN/dt=0.2$ Hz.\cite{Pon00a}  It is possible that this limit is
too conservative and could be lowered with identifiable backgrounds
and knowledge of the direction of the signal.  The width of the bands
represents the uncertainty in $<E_{\bar\nu_e}>$ due to the diffusion
approximation.\cite{pons99,Pon00a,Pon01b}  It is possible to
distinguish between stable and metastable stars, since the
luminosities when metastability is reached are always above
conservative detection limits.

\subsection{Metastable Protoneutron Stars}

\begin{figure}[htb]
\centerline{\epsfxsize=3.7in\epsfbox{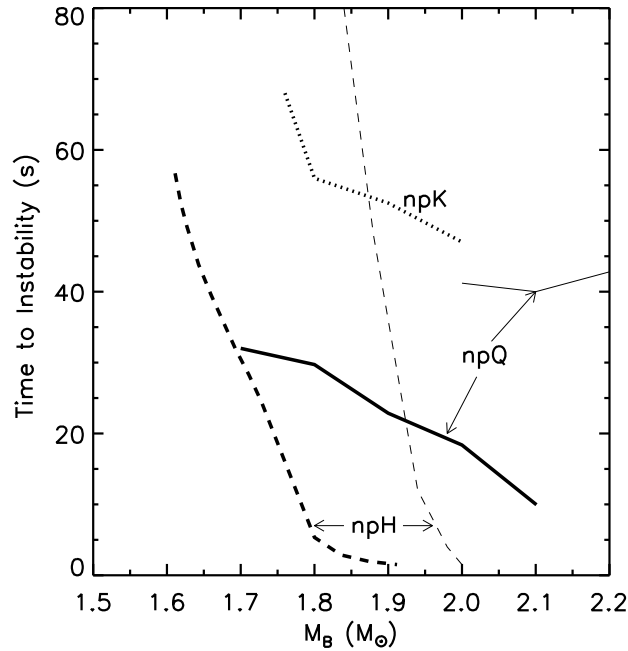}}  
\caption{Lifetimes of metastable stars versus the PNS
baryon mass $M_B$.  Thick lines denote cases in which the maximum
gravitational masses of cold, catalyzed stars are near 1.45 M$_\odot$,
which minimizes the metastability lifetimes.  The thin lines for the
$npQ$ and $npH$ cases are for EOSs with larger maximum gravitational
masses (1.85 and 1.55 M$_\odot$, respectively.)}
\label{fig:ttc}
\end{figure}

Protoneutron stars in which strangeness appears following
deleptonization can be metastable if their masses are large enough.
One interesting diagnostic that could shed light on the internal
composition of neutron stars would be the abrupt cessation of the
neutrino signal.  This would be in contrast to a normal star of
similar mass for which the signal continues to fall until it is
obscured by the background.  In Figure~\ref{fig:ttc} the lifetimes for
stars containing hyperons ($npH$), kaons ($npK$) and quarks ($npQ$)
are compared.\cite{Pon00a}  In all cases, the larger the mass, the
shorter the lifetime.  For the kaon and quark PNSs, however, the
collapse is delayed until the final stage of the Kelvin-Helmholtz
epoch, while this is not necessarily the case for hyperon-rich stars.
In addition, there is a much stronger mass dependence of the lifetimes
for the hyperon case.

Clearly, the observation of a single case of metastability, and the
determination of the metastability time alone, will not necessarily
permit one to distinguish among the various possibilities. Only if the
metastability time is less than 10--15 s, could one decide on this
basis that the star's composition was that of $npH$ matter.  However,
as in the case of SN 1987A, independent estimates of $M_B$ might be
available.\cite{THM90BB95}  In addition, the observation of two or
more metastable neutron stars might permit one to differentiate among
these models.

\newpage

\section{Neutron Stars in Their Old Age}
See the contribution from Dany Page to this symposium.

\section{Outlook}

The advent of new-generation neutrino detectors
such as Super-Kamiokande and the Sudbury Neutrino Observatory promises
thousands of neutrino events in the next Galactic supernova.  These
will provide crucial diagnostics for the supernova mechanism,
important limits on the released binding energy and the remnant mass,
and critical clues concerning the composition of high density matter.
Research in this area will ascertain the extent to which neutrino
transport is instrumental in making a supernova explode.  Other
bonuses include the elucidation of the possible role of supernovae and
neutrinos in $r-$process nucleosynthesis.

The main issues that emerge from PNS studies concern the metastability
and subsequent collapse to a black hole of a PNS containing quark
matter, or other types of matter including hyperons or a Bose
condensate, which could be observable in the $\nu$ signal.  However,
discriminating among various compositions may require more than one
such observation.  This highlights the need for breakthroughs in
lattice simulations of QCD at finite baryon density in order to
unambiguously determine the EOS of high density matter.  In the
meantime, intriguing possible extensions of supernova and PNS
simulations with $npQ$ and $npK$ matter include the consideration of
heterogenoeus structures and quark matter superfluidity.\cite{CR00}
See also the contribution from Sanjay Reddy to this symposium. 
\section{Conclusions}

\begin{figure}[htb]
\centerline{\epsfxsize=3.7in\epsfbox{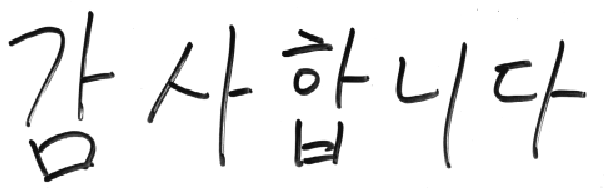}}  
\centerline{\epsfxsize=3.7in\epsfbox{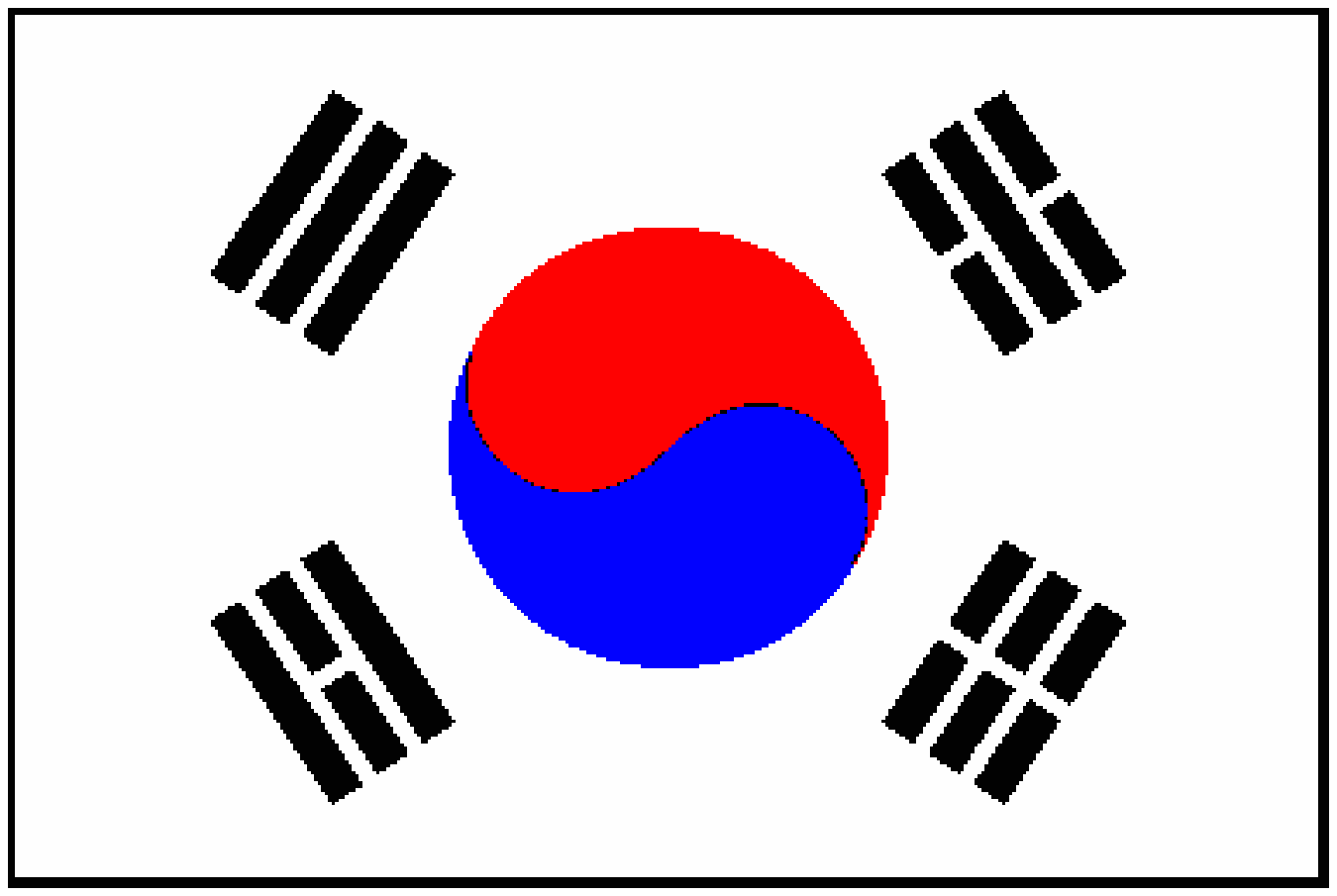}} %
%
\label{fig:thanku}
\end{figure}

\section*{Acknowledgements}

MP thanks his ``PALS'' Paul Ellis, Jim Lattimer, Jose Miralles, Dany
Page, Jose Pons, Sanjay Reddy, and Andrew Steiner for benificial
collaborations.  Research support from DOE grants FG02-88ER-40388 and
FG02-87ER-40317 (for MP, SR and SID), and NSF Grant No. 0070998 (for SID),
and travel support under the cooperative agreement DE-FC02-01ER41185
for the SciDaC project ``Shedding New Light on Exploding Stars:
Terascale Simulations of Neutrino-Driven Supernovae and Their
Nucleosynthesis'' (for MP, SR and SID) are gratefully acknowledged.

%
%
%
%

\end{document}